\documentclass{ws-procs9x6}

\usepackage{graphics}
\usepackage{psfrag}
\usepackage{epsfig}
\usepackage{psfig}
\usepackage{epsf}

\def\3{{\ss}}
\newcommand{\beq}{\begin{equation}}
\newcommand{\eeq}{\end{equation}}
\newcommand{\beqa}{\begin{eqnarray}}
\newcommand{\eeqa}{\end{eqnarray}}

\begin{document}

\author{Ulf-G. MEI{\ss}NER}

\address{Universit\"at Bonn, Helmholtz-Institut f\"ur Strahlen- und Kernphysik (Theorie) \\
Nu{\ss}allee 14-16, D-53115 Bonn, Germany\\
E-mail: meissner@itkp.uni-bonn.de} 

\title{Recent results in chiral nuclear dynamics}

\author{V\'eronique Bernard}

\address{Laboratoire de Physique Th\'eorique, Universit\'e Louis Pasteur\\
3-5, rue de l'Universit\'e, F-67084 Strasbourg, France}

\author{Evgeny EPELBAUM, 
\,\,  Walter GL\"OCKLE}

\address{Ruhr--Universit\"at Bochum, Institut f\"ur Theoretische Physik II, \\
Universit\"atsstra{\ss}e 150, D-44870 Bochum, Germany}


\maketitle

\abstracts{Some recent developments in the description of nuclear forces and 
few-nucleon dynamics
derived from chiral effective field theory are reviewed.}

\section{Introduction}
\label{intro}
Nuclear forces can be systematically analyzed in the framework of chiral effective
field theory (EFT). The starting point is a chiral Lagrangian of pions, nucleons and
external sources, like e.g. photons. At low energies, one can
expand S--matrix elements and transition currents in powers of small external 
momenta/energies and the light quark masses. In case of two (or more)
nucleons, an additional non--perturbative resummation is necessary to generate
the large S-wave scattering lengths or small nuclear binding energies. This is
most efficiently done by applying a chiral power counting scheme to the 
nucleon-nucleon (NN) potential. With that,
one solves the corresponding  Lippmann-Schwinger (LS) equation for bound
and scattering states \cite{wein}, employing a regularization procedure consistent with
the underlying symmetries. This framework allows for a systematic evaluation
of the forces between 2, 3, or 4 nucleons, the gauge--invariant coupling of
external probes or the investigation of nuclear physics in the limit of
vanishing quark masses. In what follows, I will briefly address some of these
issues. Space forbids to account for all the interesting work done by many
researchers and references are only given that are of direct relevance to the
topic under consideration.

\section{Few--nucleon systems}
\label{sec:1}
\begin{figure}[t]
{\psfig{file=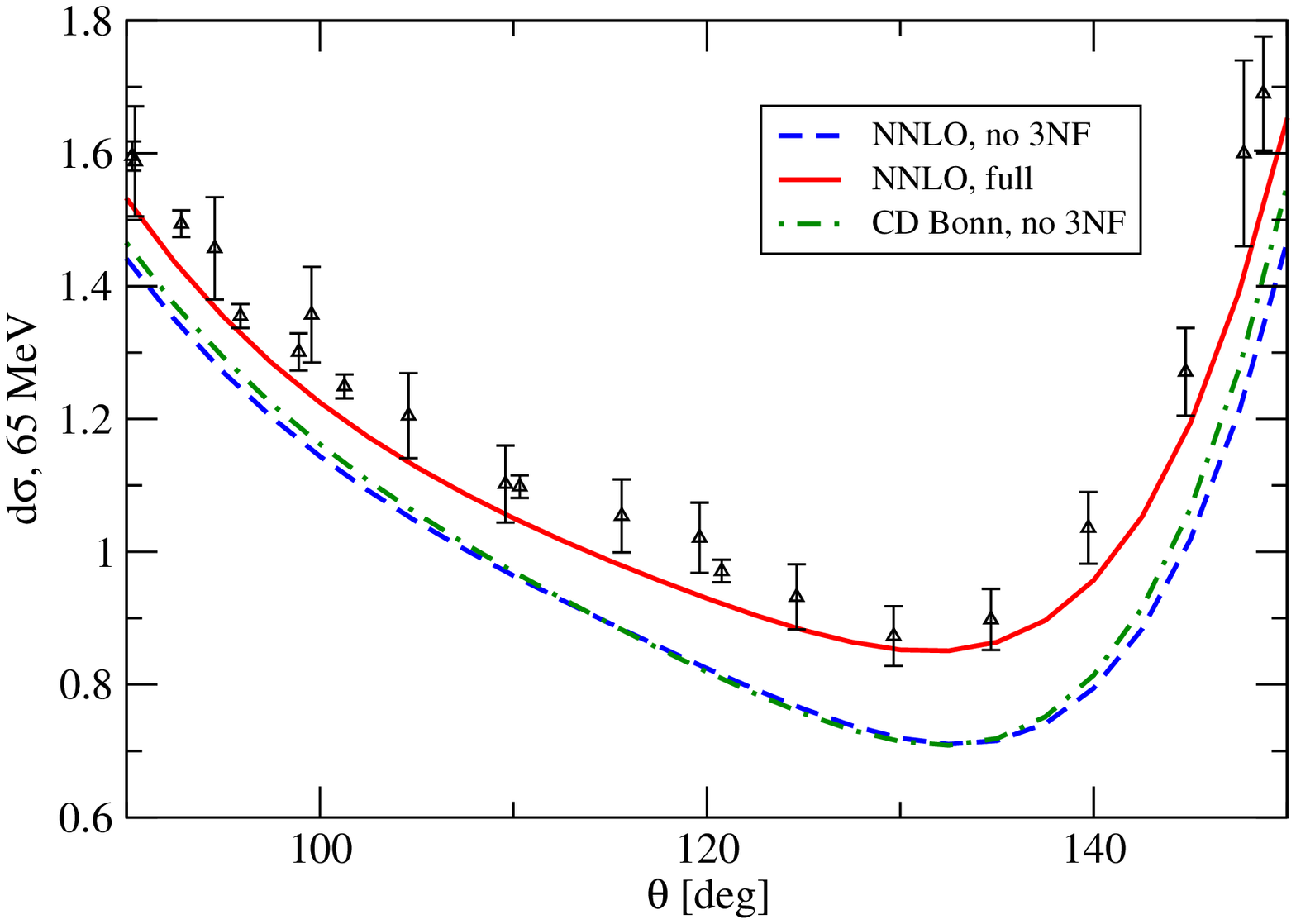,width=5.3cm}\hfill
\psfig{file=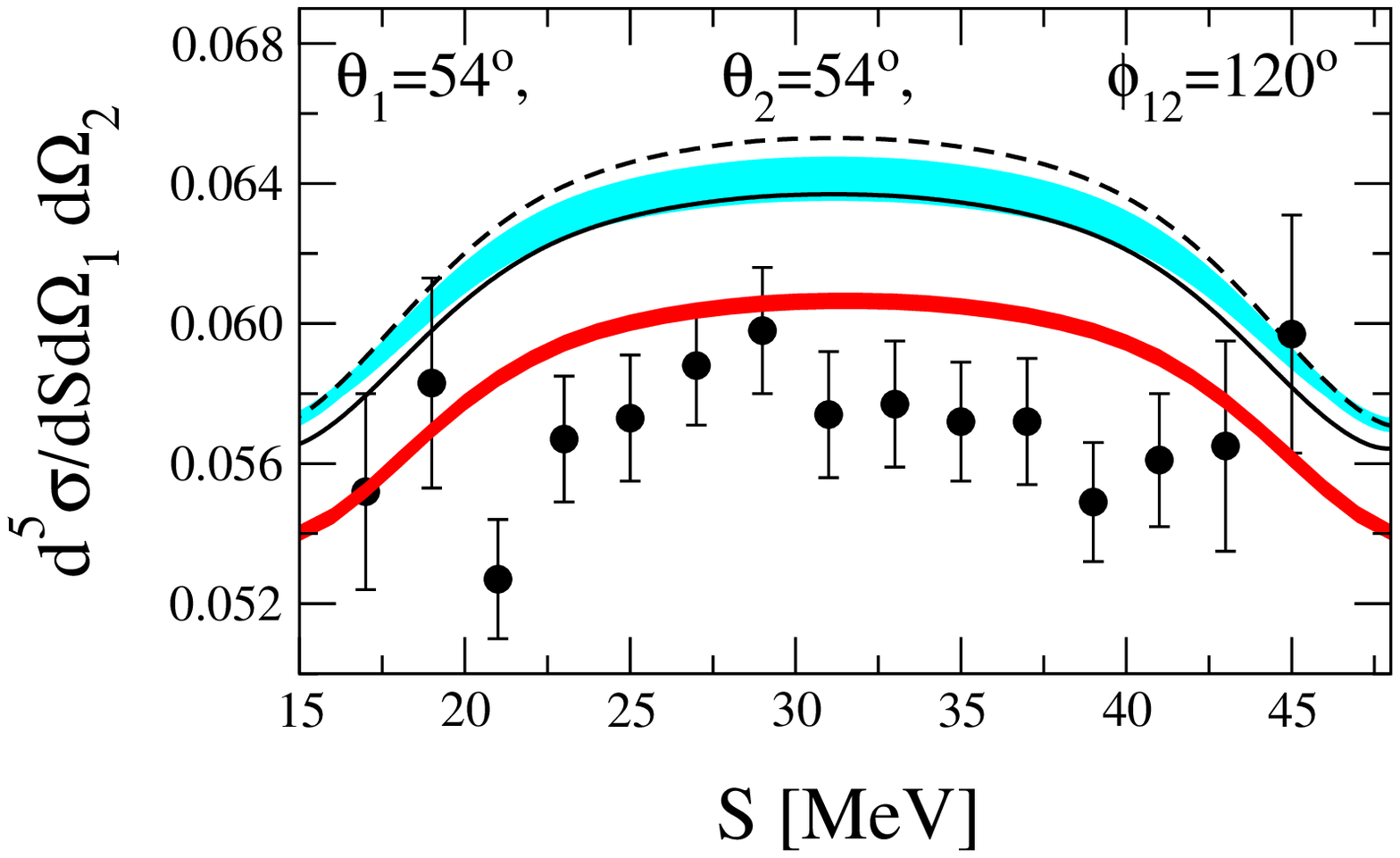,width=6.cm}  }
\caption[fig1]{\label{fig1} Left: Differential cross section for $nd$ scattering at
65 MeV. Solid (dashed) line: NNLO prediction with (without) 3NF. The dot-dashed line
gives the prediction based on the CD-Bonn potential without 3NF. 
Right: $pd$ breakup cross section data in [mb MeV$^{-1}$ sr$^{-2}$]
along the kinematical locus S (in MeV) at 65 MeV. The data are from \cite{zejma}.
NNLO predictions (dark shaded band) compared to the conventional 
NN forces$+$TM 3NF predictions (light shaded band). The solid (dashed) line refers to the 
AV18$+$URBANA IX (CD Bonn$+$TM') results. }
\end{figure}

The two-nucleon force can be well described within this framework, see
e.g. \cite{EGM2}. More interesting is the three-nucleon force (3NF), which has
been one of the most sought after objects in nuclear physics. In chiral EFT, it
first appears at next-to-leading order (NLO), but these contributions can be
shown to vanish \cite{wein,bira,EGM1}. At NNLO, there are three different
topologies: two--pion--exchange between 3 nucleon lines, one--pion--exchange 
between a 4N contact term and a nucleon line and a 6N contact interaction. Two 
low--energy constants (LECs) (related to the latter two topologies)
appear. These can be fixed from the triton binding energy (BE) and the $nd$ 
doublet scattering length. With that, the 3NF is completely determined and the
properties of few-nucleon systems can be calculated in a parameter--free
manner \cite{Eprc}. In most cases, the predictions based on chiral EFT are
very close to the ones based on more conventional approaches, that is the
so-called high precision NN potentials combined with some (non-chiral) 3NF, see e.g. the
differential cross section for elastic $nd$ scattering at 65 MeV (left panel
of Fig.~\ref{fig1}). 
\noindent 
There are, however, exceptions, see the specific break-up
configuration in the right panel of Fig.~\ref{fig1}. This is indeed the break-up
observable where the largest deviations of the two approaches are observed.  
Solving the pertinent Faddeev-Yakubovsky equations for the 4N system, one can
calculate e.g. the BE of $^4$He. We find \cite{Eprc}
\beq
{\rm BE(^4He)} = 29.51 \ldots 29.98~{\rm MeV} \quad {\rm for}~~\Lambda = 500
\ldots 600~{\rm MeV}~,
\eeq
where $\Lambda$ is the cut-off in the regularized LS equation. The empirical
value (corrected for isospin breaking effects) is $29.8\pm 0.1$~MeV. Thus there
is very little room for a 4NF. Interestingly, the first corrections to the chiral 3NF 
are free of 6N contact 
LECs and are presently worked out.

\section{Nuclear forces in the chiral limit}
\label{sec:2}
Because of the smallness of the up and down quark masses, one does not expect
significant changes in systems of pions or pions and one nucleon when the
quark masses are set to zero (with the exception of well understood chiral
singularities like e.g. in the pion radius or the nucleon
polarizabilities). The situation is more complicated for systems of two (or
more) nucleons, as first discussed in EFT in \cite{BBSvK}. Here, I report on
similar work \cite{EMG} that is mostly concerned with the properties of the
deuteron and the S-wave scattering lengths as a function of the quark (pion)
mass. These questions are not only of academic interest, but also of practical
use for interpolating results from lattice gauge theory. E.g. the S-wave
scattering lengths have been calculated on the lattice using the quenched 
approximation \cite{Fukug95}. Another interesting application is related to
imposing bounds on the time-dependence of some fundamental coupling constants
from the NN sector, as discussed in \cite{Beane02}.
At NLO the following contributions have to be
accounted for (in addition to the LO OPE and contact terms without derivatives): 
i) contact terms with two derivatives or one $M_\pi^2$--insertion, 
ii) renormalization of the OPE,
iii) renormalization of the contact terms, and
iv) two--pion exchange (TPE).
This induces {\em explicit} and  {\em implicit} quark mass dependences. In the
first category fall the pion propagator that becomes Coulomb-like in the
chiral limit or the $M_\pi^2$ corrections to the leading contact terms. These
are parameterized by the LECs $\bar D_{S,T}$ at NLO. These LECs can at present 
only be estimated using dimensional analysis and resonance saturation \cite{EGME}. 
The implicit pion mass dependence enters at NLO through the pion--nucleon
coupling constant,\footnote{Note that the quark mass dependence of the nucleon
mass only enters at NNLO.} expressed through the pion mass dependence of $g_A/F_\pi$,
\beq
\label{deltaCL}
\Delta = \left( \frac{g_A^2 }{16 \pi^2 F_\pi^2} - \frac{4 }{g_A}
\bar{d}_{16} + \frac{1}{16 \pi^2 F_\pi^2} \bar{l}_4 \right) (M_\pi^2 - \tilde M_\pi^2) - 
\frac{g_A^2 \tilde M_\pi^2}{4 \pi^2 F_\pi^2} \ln \frac{\tilde M_\pi}{M_\pi} \, .
\eeq
Here $\bar l_4$, $\bar d_{18}$ and $\bar d_{16}$ are LECs related to 
pion and pion--nucleon interactions, and the value of the pion mass is denoted
by $\tilde M_\pi$ in order to distinguish it from the physical one denoted by
$M_\pi$. In particular, $\bar d_{16}$ has been determined in various fits to
describe $\pi N \to \pi\pi N$ data, see \cite{FBM}.

\begin{figure}[t]
\centerline{
\psfig{file=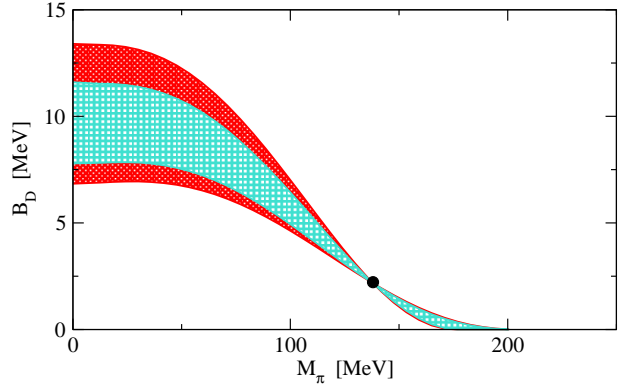,width=9cm}}
\caption[fig2]{\label{fig2} Deuteron BE versus $\tilde M_\pi$. The shaded areas show 
allowed values. The light shaded band corresponds to our main result with  
the uncertainty due to the unknown LECs $\bar D_{S,T}$.
The dark shaded band gives the additional uncertainty due to the uncertainty of $\bar d_{16}$.
The heavy dot shows the BE for the physical case $\tilde M_\pi = M_\pi$.}
\end{figure}

The deuteron BE as a function of the pion mass is shown in Fig.\ref{fig2},
we find that the deuteron is stronger bound in the chiral limit than in the
real world with the BE  $B_{\rm D}^{\rm CL} =  9.6 \pm 1.9 {{+ 1.8} \atop  {-1.0}}$ MeV,
where the the  
first indicated error refers to the uncertainty in the value of $\bar D_{^3S_1}$ and 
$\bar d_{16}$ being set to its average value 
while the second indicated error shows the additional uncertainty due to the uncertainty
in the determination of $\bar d_{16}$
\noindent We find no other bound states, although the higher S=1 partial waves
rise linear with momentum due to the Coulomb-like pion propagator. Other
deuteron properties are given in the table. We note in particular the large
uncertainty for the quadrupole moment, which is related to its increased
sensitivity to short distance physics. 
\begin{table}[h]
\tbl{Deuteron properties at NLO compared to the data.\vspace*{1pt}}
{\footnotesize
\begin{tabular}{|c|c|c|c|}
\hline
{} & {} &  {} & {}\\[-1.5ex]
{} &  physical case   &  chiral limit & experiment\\[1ex]
\hline
{} &{} &{} &{}\\[-1.5ex]
$B_{\rm D}$ [MeV]                & $2.17$  & $9.6 \pm 1.9 {{+ 1.8} \atop  {-1.0}}$ &
2.22456612(12) \\[1ex]
$Q_{\rm D}$ [fm$^2$]             & $0.274$ & $0.247 \pm 0.030 {{ + 0.008} \atop {-0.005}} $
& 0.2859(3)\\[1ex]
$r_{\rm D}$ [fm]                 & $1.975$ & $1.266 \pm 0.085 {{ + 0.044} \atop {-0.034}}$
& 1.9671(6) \\[1ex]
$\mu_{\rm D}$ [n.m.]             & $0.860$ & $0.820 \pm 0.002 {{ + 0.003} \atop {-0.006}}$
& 0.8574382284(94)\\[1ex] 
\hline
\end{tabular}\label{tab1} }
\end{table}
Last but not least,
we found smaller (in magnitude) and more natural values for the two 
S--wave scattering lengths in the chiral limit,
$a_{\rm CL} (^1S_0) = -4.1 \pm 1.6 {{+ 0.0} \atop 
{-0.4}} \,{\rm fm}$,  and
$a_{\rm CL} (^3S_1) = 1.5 \pm 0.4 {{+ 0.2} \atop \\ 
{ -0.3}}\, {\rm fm}\,$.
As stressed in \cite{EMG}, one needs lattice data for pion masses below
200 MeV to perform a stable interpolation to the physical value of $M_\pi$.
We conclude that nuclear physics in the chiral limit is much more natural than 
in the real world.

\section{Pion--deuteron scattering}
\begin{figure}[t]
\centerline{
\psfig{file=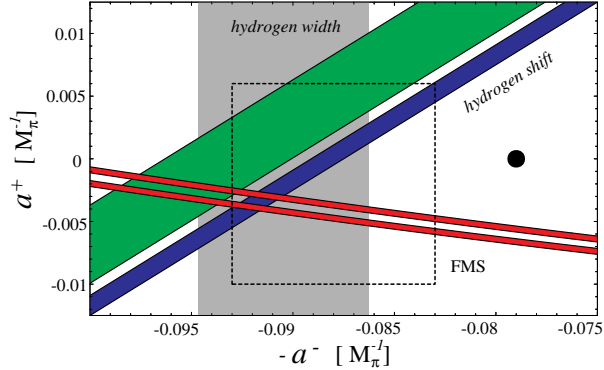,width=8cm}}
\caption[fig3]{\label{fig3}
    Plot of $a^+$
    vs $-a^-$. The light shaded region and the dark band are from the experimental
    pionic-hydrogen width and shift, respectively, taken from Ref.~\cite{Schroder:uq}.
    The shaded region above the dark band is the hydrogen shift computed in
    bound state EFT~\protect\cite{Bern}.
    The dotted line encompasses the constraints from $\pi$-N phase shift data
    and is taken from Ref.~\protect\cite{FMS}.
    The dot is leading order $\chi$PT (current algebra). The two parallel bands
are from
    Eq.~(\ref{thefullthing}) evaluated with the NLO wavefunction with an
    ultraviolet cutoff of $400$~MeV (upper curve) and $600$~MeV (lower curve).
}
\end{figure}

The pion-nucleon ($\pi$N) S-wave scattering lengths are quantities of
fundamental importance in hadronic physics. They provide an
important test of QCD symmetries and the pattern of their breaking.
They also provide a crucial constraint on the $\pi$N
interaction and, as such, have an impact on our understanding of
nucleon-nucleon (NN) scattering, the three-nucleon force, and
pion-nucleus scattering. They can be e.g. determined by analyzing low
energy scattering data making use of chiral perturbation theory, for the
most recent analysis in the isospin limit see \cite{FM4}. However, a 
novel evaluation of electromagnetic corrections has cast some doubt
on the existing phase shift analyses \cite{FMiso}.
An independent approach to the $\pi$N scattering lengths involves
analyzing pionic-atom level shifts and widths. In the Coulombic
$\pi^-$-p system, the strong-interaction shift in the energy (the width)
can be used to infer a value for $a^- + a^+$ ($a^-$). Here,  $a^-$ ($a^+$)
refers to the isovector (isoscalar) scattering length.
Thus, neither $\pi$N scattering 
nor the $\pi^-$-p atom provide a
strong constraint on $a^+$. The isoscalar scattering length may well
be probed more directly in the $\pi^-$-d atom. 
The most recent measurement~\cite{Hauser:1998yd}
of the $\pi$-d atomic-level shift yields:
$a_{\pi d}=(-0.0261 \pm 0.0005) \;+\; i\; (0.0063 \pm 0.0007)\; {M_\pi^{-1}}$
for the $\pi$-d scattering length -- a measurement that is
remarkably accurate given the usual level of precision in
hadronic-physics experiments. In order to precisely relate this number to
$a^+$ an EFT of threshold $\pi$-d scattering is required. In \cite{BBEMP}
we have formulated a novel power counting which accounts for the fact that
the deuteron binding momentum $\gamma \simeq 45\,$MeV is sizeably smaller than the
pion mass \cite{BS}. 
This lets one isolate easily the dominant contributions and calculate the
corrections. To fourth order, one gets for the $\pi$-d scattering length,
\begin{eqnarray}
{\rm Re}\,a_{\pi d} =2\frac{(1+\mu)}{(1+\mu /2)}\,\left( a^+ \, +\, 
(1+\mu )\Big\lbrack (a^+)^2-2(a^-)^2 \Big\rbrack 
\frac{1}{2{\pi^2}}\Bigg\langle 
\frac{1}{{\vec q\,}^{\, 2}}\Bigg\rangle_{\rm wf}\right. && \nonumber\\
\left.  \quad
+(1+\mu )^2\Big\lbrack (a^+)^3-2(a^-)^2(a^+-a^-) \Big\rbrack 
\frac{1}{4{\pi}}\Bigg\langle 
\frac{1}{|{\vec q}\, |}\Bigg\rangle_{\rm wf}\, \right) 
\,+\, {a_{(boost)}}\,,&&
\label{thefullthing}
\end{eqnarray}
where some higher-order effects are subsumed into the  $\pi$N scattering lengths,
$\mu = M_\pi/m_p \simeq 1/7$ is the small threshold parameter,  the brackets
$\langle~\rangle_{\rm wf}$ denote a deuteron matrix--element in momentum space,
and the boost correction is discussed in \cite{BBEMP}. The
resulting relationship between $a^-$ and $a^+$ is displayed in
Fig.~\ref{fig3}.
Our model-independent extraction of the $\pi$N scattering lengths results from
the overlap region in Fig.~\ref{fig3} using the hydrogen shift computed
in bound state EFT~\cite{Bern} (upper curve). We find
\beq
a^{-} = 0.0936 \pm 0.0011\; {M_\pi^{-1}}\,  , \,\,\,
a^{+} = -0.0029 \pm 0.0009\; {M_\pi^{-1}}\ .
\eeq
If one uses instead the model-dependent extraction using the hydrogen shift of
Ref.~\cite{Schroder:uq}, these numbers change to $a^{-}= 0.0917 \pm 0.0013 \, {M
_\pi^{-1}}$,
$a^{+}= -0.0034 \pm 0.0008\; {M_\pi^{-1}}\,$. We note that our result is consistent with
the recent work of Ref.~\cite{Ericson:2000md}. What needs to be done is a systematic
inclusion of isospin--breaking corrections in the EFT for the $\pi$-d system.

\section{Neutral pion electroproduction off deuterium}
\label{summ}
\begin{figure}[t]
\centerline{
\psfig{file=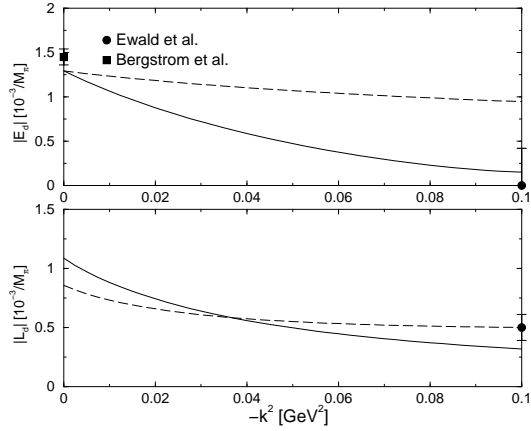,height=5.8cm}}
\caption[fig4]{\label{fig4}
   Modulus of the threshold S-wave multipoles $|E_d|$ and $|L_d|$ as a function of the
   photon virtuality in comparison to the photon point data
   from SAL \protect\cite{SALd} and the electroproduction data from
   MAMI  \protect\cite{Ewald}. The sign of the experimental result
   for $L_d$ is taken to agree with the theoretical prediction.
   Solid (dashed) lines: Fit~2 (1), as explained in \protect\cite{KBM}. 
}
\end{figure}

The last topic I would like to discuss briefly is neutral pion
electroproduction off deuterium. While $\pi^0$ production off protons
has become one of the major testing grounds of chiral $\pi$N dynamics,
the elusive neutron photo/electroproduction amplitude can only be obtained 
from experiments on light nuclei. While the photoproduction case $\gamma d
\to\pi^0 d$ has been quite a success (the chiral prediction of \cite{BBLMvK}
was verified by the SAL experiment \cite{SALd}), the first measurement of pion
electroproduction off deuterium at MAMI~\cite{Ewald} was in stark disagreement with
the third order threshold prediction of~\cite{BKM}, in which the single
scattering contribution was scaled to the value at the real photon point.
This was significantly improved in the above threshold (partial) fourth order
calculation of
\cite{KBM}, where the two appearing LECs were determined in two different ways.
In addition, we also used deuteron wave functions derived from chiral nuclear EFT.
With that, the existing differential cross section data at four different
excess energies and three photon polarizations each could be satisfactorily
described. Furthermore, the dependence of the transverse and the longitudinal
S-wave amplitude on the photon virtuality is now in agreement with the data,
see Fig.\ref{fig4} (although some discrepancies remain). The two lines shown in the
figure can be considered as a measure of the theoretical uncertainty at this order.
Clearly, the calculation presented in \cite{KBM} needs to be improved, in particular,
the fourth order corrections to the P--waves and the three-body terms
have to be included (note that similar work for the P--waves in neutral
pion production off protons has only appeared recently \cite{BKMa}). This reaction shows nicely
the intricate interplay of chiral nucleon and nuclear dynamics.

\section*{Acknowledgements}
I am grateful to Koichi Yamawaki for inviting me and to all the organizers for their
magnificent job. My collaborators Silas Beane, 
Hiroyuki Kamada,
Hermann Krebs, Andreas Nogga, Daniel Phillips and  
Henryk Witala are thanked for sharing their
insight into the various topics discussed here.


\begin{thebibliography}{}
%
%
\bibitem{wein} S.~Weinberg, Nucl. Phys. \textbf{B363}, 3 (1991).
\bibitem{EGM2} E.~Epelbaum, W.~Gl\"ockle and  Ulf--G.~Mei\3ner, 
Nucl. Phys.  \textbf{A671}, 295 (2000).
\bibitem{bira} U. van Kolck, Phys. Rev. \textbf{C49}, 2932 (1994).
\bibitem{EGM1} E.~Epelbaoum, W.~Gl\"ockle and  Ulf--G.~Mei\3ner, 
Nucl. Phys. \textbf{A637}, 107 (1998).
\bibitem{Eprc} E.~Epelbaum, {\it et al.}, 
Phys. Rev.  \textbf{C66}, 064001 (2002).
\bibitem{zejma} J.~Zejma et al., Phys. Rev. \textbf{C55}, 42 (1997).
\bibitem{BBSvK}  S.R.~Beane, et al., Nucl. Phys. \textbf{A700}, 377 (2002).
\bibitem{EMG} E.~Epelbaum, Ulf--G.~Mei\3ner, and W.~Gl\"ockle,
 Nucl. Phys.  {\bf A714}, 535 (2003); nucl-th/0208040 
\bibitem{Fukug95} M.~Fukugita, {\it et al.}, Phys. Rev. \textbf{D52}, 3003 (1995).
\bibitem{Beane02} S.R.~Beane and M.J.~Savage, Nucl. Phys. {\bf A713}, 148 (2003).
\bibitem{EGME}  E.~Epelbaum, Ulf--G.~Mei\3ner, W.~Gl\"ockle, and Ch.~Elster, 
Phys. Rev. \textbf{C65}, 044001 (2002).
\bibitem{FBM} N. Fettes, V. Bernard, and Ulf--G.~Mei\3ner,
Nucl. Phys. \textbf{A699}, 269 (2000);
N.~Fettes, doctoral thesis, published in 
{\it Berichte des Forschungszentrum J\"ulich}, 
\textbf{3814}, (2000).
\bibitem{FM4}
N.~Fettes and Ulf-G.~Mei{\ss}ner, Nucl. Phys. {\bf A676}, 311 (2000).
\bibitem{FMiso}
N.~Fettes and Ulf-G.~Mei{\ss}ner, Nucl. Phys. {\bf A693}, 693 (2001).
\bibitem{Hauser:1998yd}
P.~Hauser {\it et al.}, Phys. Rev.  {\bf C58}, 1869 (1998).
\bibitem{BBEMP}S.R. Beane {\it et al.}, hep-ph/0206219.
\bibitem{BS}S.R. Beane and M.J.~Savage, nucl-th/0204046.
\bibitem{Bern} J.~Gasser {\it et al.},  Eur. Phys. J. {\bf C26}, 13 (2002).
\bibitem{Schroder:uq}
H.C.~Schr\"oder {\it et al.},
Phys.\ Lett.\  {\bf B469}, 25 (1999);
Eur.\ Phys.\ J.\  {\bf C21}, 473 (2001).
\bibitem{FMS}N.~Fettes, Ulf-G.~Mei{\ss}ner, and S.~Steininger, 
Nucl.\ Phys.\  {\bf A640}, 199 (1998).
\bibitem{Ericson:2000md}
T.E.~Ericson, B.~Loiseau and A.W.~Thomas,
Phys.\ Rev.\ {\bf C66}, 014005 (2002).
\bibitem{BBLMvK}
S.~R.~Beane, V.~Bernard, T.-S.H.~Lee, Ulf-G.~Mei{\ss}ner 
and U.~van Kolck, Nucl. Phys. {\bf A618}, 381 (1997).
\bibitem{SALd} J.C.~Bergstrom, {\it  et al.}, Phys. Rev.{\bf  C57}, 3203 (1998).
\bibitem{Ewald} I.~Ewald {\it et al.}, Phys.\ Lett.\ {\bf B499},238 (2001).
\bibitem{BKM}V.~Bernard, H.~Krebs and Ulf-G.~Mei{\ss}ner,
Phys.\ Rev.\ {\bf C61}, 058021 (2000).
\bibitem{KBM}H. Krebs, V. Bernard, and  Ulf-G.~Mei{\ss}ner,
Nucl. Phys. {\bf A713}, 405 (2003).
\bibitem{BKMa}
V.~Bernard, N.~Kaiser and Ulf-G.~Mei{\ss}ner, Eur.\ Phys.\ J.\ {\bf A11}, 209 (2001).
\end{thebibliography}
\end{document}